\documentclass{article}
\usepackage{graphicx} 
\usepackage{amsmath}  
\usepackage{authblk}  
\usepackage{natbib}   
\usepackage{amsfonts} 
\usepackage{amssymb}  
\usepackage{colortbl}
\usepackage{xcolor}
\usepackage{array}
\usepackage{booktabs}
\DeclareUnicodeCharacter{03BA}{$\kappa$}

\title{\textbf{AI-Driven Hybrid Ecological Model for Predicting Oncolytic Viral Therapy Dynamics}}
\date{}
\author[1]{Abicumaran Uthmacumaran}
\author[2]{Juri Kiyokawa}
\author[3]{Hiroaki Wakimoto\footnote{Corresponding author: \texttt{hwakimoto@mgh.harvard.edu}}}
\affil[1]{Dept. Experimental Surgery, McGill University, Montreal, Canada}
\affil[2]{Department of Neurosurgery, Institute of Science Tokyo, Tokyo, Tokyo, Japan}
\affil[3]{Department of Neurosurgery, Massachusetts General Hospital, Harvard Medical School, Boston, USA}

\begin{document}

\maketitle

\section{Abstract}

 Oncolytic viral therapy (OVT) is an emerging precision therapy for aggressive and recurrent cancers. However, its clinical efficacy is hindered by the complexity of tumor-virus-immune interactions and the lack of predictive models for personalized treatment. This study develops a data-driven, and AI-powered computational model combining time-delayed Generalized Lotka-Volterra (GLV) equations with advanced optimization algorithms, including Genetic Algorithms (GA), Differential Evolution (DE), and Reinforcement Learning (RL) to optimize OVT oscillations' growth, and damping. We hypothesize that the model can provide accurate, real-time predictions of OVT responses while identifying key biomarkers to enhance therapeutic efficacy. We demonstrate the model’s strong predictive accuracy (MSE \textless 0.02, $R^2$ \textgreater 0.82) and its capacity to identify experimentally validated biomarkers such as TNF, NF-kB, CD81, TRAF2, IL18, and BID, among other inflammatory cytokines and extracellular matrix reconstruction factors, despite being causally agnostic and unaware of specific experimental conditions or therapeutic combinations. Gene set enrichment analysis confirmed these biosignatures as critical predictors of tumor progression and indicated that photodynamic therapy activates immune responses similar to those elicited by combined OVT and immune checkpoint inhibitors. This hybrid model represents a significant step toward precision oncology and computational medicine, enabling longitudinal, adaptive treatment regimens, and the development of targeted immunotherapies based on molecular signatures, potentially improving patient outcomes.

\section{Introduction}

Oncolytic viral therapy (OVT) presents a promising frontier in cancer immunotherapy, repurposing genetically engineered viruses to selectively target and destroy tumor cells while sparing normal tissue \citep{Lin2023, Guo2014}. Despite its potential in the treatment of aggressive and recurrent tumors, OVT's clinical translation faces significant challenges, including the multiscale complexity of tumor-virus-immune interactions and the difficulty of tailoring therapies to personalized tumor microenvironments\citep{Kiyokawa2019}. One major knowledge gap is the absence of predictive computational or mathematical models that can accurately simulate these complex interactions and provide real-time feedback for optimizing therapies. Without such predictive tools, clinicians are limited in their ability to forecast OVT progression and personalize treatment regimens, leading to suboptimal efficacy and patient outcomes\citep{alzahrani2019multiscale}.

As such, this study develops a proof-of-concept data-driven predictive model to optimize OVT responses, with the potential to translate toward patient-centered care. The artificial intelligence (AI)-powered hybrid model combines a predator-prey model with advanced algorithms and feature salience analysis to decode the biological underpinnings of tumor-virus-immune dynamics, making it explainable rather than a "black box" approach. Delay parameters, growth, and decay rates provide clinically meaningful predictors, correlating temporal features and oscillatory patterns in OVT-tumor dynamics with therapy progression (response), enabling adaptive, patient-specific treatment strategies. To our understanding, this is the first explainable AI-integrated, adaptive algorithm to model OVT dynamics from an ecological and dynamical systems perspective, enabling continuous monitoring and personalized treatment adjustments (for dosing). 

\subsection*{Insights from Oncolytic Adenovirus Therapy and Open Questions on Predictive Modelling}

The study by \citet{Kiyokawa2021} investigated the synergistic effects of ICOVIR17, a hyaluronidase-expressing oncolytic adenovirus, on a murine glioblastoma model (005 cells) treatment in combination with anti-PD-1 immune checkpoint blockade. Key findings of the combination therapy included an increased ratio of CD8+ T cells to regulatory T cells and enhanced cytotoxicity of T cells against GBM cells ex vivo \citep{Kiyokawa2021}. The underlying mechanism was found to be that ICOVIR17 degrades hyaluronan (HA) in the extracellular matrix (ECM), thus, enhancing tumor-infiltrating immune cells, activating NF-kB signaling in macrophages, and promoting a proinflammatory tumor microenvironment. These changes were associated with increased expression of PD-L1 and an enhancement of therapeutic efficacy when combined with PD-L1 blockade.

However, some open questions from these findings include: How do the virus and anti-PD-1 therapies interact with each other? Can we identify molecular mediators that drive such interactions and complex dynamics? Can a mathematical or computational model accurately predict the contributions of tumor microenvironment (ECM) remodeling and immune activation to therapeutic efficacy, including the dynamics of immune markers such as TNF and NF-kB identified in this study?

To address these questions, we will use the data from \citet{Kiyokawa2021} to train and inform our computational modeling, particularly in exploring the interactions between oncolytic virotherapy and immune checkpoint inhibitors.

\subsection{Oncolytic Viral Therapy (OVT) as an Ecological Dynamics Model}

State-of-the-art OVT models in mathematical oncology and computational medicine, ranging from ordinary differential equations (ODEs) to reaction-diffusion equations (RDEs), to attractor reconstruction, have explored tumor-virus interactions, capturing complex dynamics like collective oscillations and chaos \citep{Sherlock2023}, \citep{Wang2019}, \citep{Heidbuechel2020}, \citep{Ramaj2023}, \citep{Jenner2022}. Early ODE models incorporated tumor-virus heterogeneity, capturing dormancy and recurrence, though they struggled with parameter optimization and data availability \citep{Karev2006}. RDE-based multiscale models simulated virus spread and tumor responses, revealing oscillatory dynamics but were sensitive to stochastic fluctuations \citep{Paiva2009}. Nonlinear ODE models showed complex oscillatory behaviors, including limit cycles and chaotic attractors, highlighting time delay and immune-tumor interactions, though they lacked empirical validation \citep{Agarwal2011}. Recent ODE models with viral and tumor cell densities, optimized to experimental data (\(R^2 = 0.4286\)), had limitations in modeling virus replication dynamics \citep{Jenner2018}. 

\citet{Wodarz2009} developed an ODE model emphasizing spatial constraints, but it lacked immune response and molecular insights, limiting its clinical predictive power. \citet{Mahasa2017} used Delayed-Differential Equations (DDEs) to show that partial normal-cell infection enhances viral potency, but with high parameter sensitivity and no clinical validation. Similarly, \citet{Friedman2018} modeled combination therapy using RDEs, noting reduced efficacy with higher inhibitor doses, but lacked molecular mechanisms. Agent-based models (ABMs) have also been explored, such as the glioblastoma model by \citet{Jenner2022}, which showed that stromal density impacts therapeutic efficacy, but lacked long-term clinical validation and predictive ability for adaptive dynamics. 

Current OVT models, while informative, face significant challenges in clinical translatability, including limitations in parameter optimization, patient-specific dynamics, and forecasting clinically relevant biomarkers. These challenges highlight the potential of DDEs and RDEs to predict complex attractors and oscillatory dynamics but also underscore the need for enhanced clinical data-based validation and predictive robustness. In principle, the complex oscillatory dynamics of OVT ecosystems—such as limit cycles and chaotic attractors—enable adaptive timing and dosage strategies to maintain tumor-virus dynamics in a non-therapy-resistant oscillatory phase. This resembles predator-prey systems, where oscillations prevent bifurcations into resistant or eradicated states. In this context, OVT can steer malignant cells in a feedback loop between treatment-resistant and therapy-susceptible (responding) phases, adapting dynamically to therapeutic interventions. 

We predict that multimodal therapies, like OVT combined with immune checkpoint inhibitors or photodynamic therapy with modulated light pulses, informed by predictive biomarkers identified by our model, can reinforce these oscillations by synchronizing treatment (entrainment) with the system's natural periodicity, thus maintaining susceptibility while avoiding resistance \citep{Agarwal2011, Shimizu2023, Kiyokawa2021}.  Given this, we hypothesize that combining DDEs with the Generalized Lotka-Volterra (GLV) model, a predator-prey framework, provides an optimal basis to capture the evolutionary dynamics, temporal delays, and feedback loops critical to virus-tumor microenvironment interactions. Additionally, incorporating AI algorithms to optimize hyperparameter tuning could enhance predictive power and replicate experimental data more efficiently.

\section{Methods}

\subsection{Generalized Lotka-Volterra Equations with Delay}

 The GLV equations with time delay best-fitting the tumor-immune-virus population dynamics are given by:

\[
\frac{dx(t)}{dt} = a \cdot x(t) \left( 1 - \frac{x(t)}{K_x + \text{growth\_mod\_x} \cdot y(t)} \right) - b \cdot x(t) \cdot y(t - \text{delay}) - \text{damping}
\]

\[
\frac{dy(t)}{dt} = -c \cdot y(t) \left( 1 - \frac{y(t)}{K_y + \text{growth\_mod\_y} \cdot x(t)} \right) + d \cdot x(t - \text{delay}) \cdot y(t) - \text{damping}
\]

\begin{itemize}
    \item \( x(t) \): Population of tumor cells (prey) at time \( t \).
    \item \( y(t) \): Population of virus-infected cells (predator) at time \( t \).
    \item \( a \): Intrinsic growth rate of tumor cells.
    \item \( b \): Rate at which tumor cells are killed by virus-infected cells.
    \item \( c \): Intrinsic death rate of virus-infected cells.
    \item \( d \): Rate at which virus-infected cells are produced by interactions with tumor cells.
    \item $K_x$: Carrying capacity of the tumor cell population.
    \item $K_y$: Carrying capacity of the virus-infected cell population.
    \item \( \text{growth\_mod\_x} \): Modifier for the growth rate of tumor cells influenced by the virus-infected cell population.
    \item \( \text{growth\_mod\_y} \): Modifier for the growth rate of virus-infected cells influenced by the tumor cell population.
    \item \( \text{damping\_x} \): Damping coefficient for the tumor cell population, representing self-limiting growth or other inhibitory effects.
    \item \( \text{damping\_y} \): Damping coefficient for the virus-infected cell population.
    \item \( \text{delay} \): Time delay in the interaction between the two populations, representing the time lag between infection of tumor cells by the virus and the resultant effect on the population sizes.
\end{itemize}

The introduction of time delay terms allows the nonlinear model to account for past states when modeling the current interaction dynamics between tumor cells, infected cells, and viruses over time\citep{aeedian2022effect, elsadany2018dynamical, akjouj2024complex, hernandez1997lotka}.  Thus, the time-delay GLV  model predicts self-organizing, higher-level emergence of multiscale ecological behaviors in tumor-immune-virus interactions within heterogeneous microenvironments.

\subsection{Computational Algorithms for Oncolytic Viral Therapy (OVT) Dynamics}

Advanced computational algorithms were used to analyze OVT dynamics datasets, treating the interaction between tumor cells (prey) and virus-infected cells (predators) using the time-delay GLV equations. The GLV model is enhanced with dynamic growth rates, adaptive learning in their competitive tumor-immune environmental interactions, damping effects, and delay terms to simulate the complex interactions observed in the complex experimental data\citep{aeedian2022effect,  akjouj2024complex}. Other stochastic and partial differential equations including reaction-diffusion systems and general time-delay differentials were first screened before settling on the GLV as the optimal model by use of statistical performance metrics (e.g., R-square, MSE fit, etc.).

The GLV model captures the predator-prey dynamics between tumor and virus-infected cells, incorporating terms for dynamic growth rates, self-limiting behavior, and interaction delays. The delay term reflects the time lag between viral infection and the subsequent effect on tumor viability. The goal is to accurately simulate the oscillatory behaviors observed in the experimental data in representing the evolutionary dynamics of OVT interactions with the tumor ecosystems.

By optimizing the GLV model parameters using Genetic Algorithms (GA) and Differential Evolution (DE), we minimized the mean squared error (MSE) between the hybrid computational model's predictions and the experimental data. These algorithms can dynamically adjust the damping and growth rates of the model. Reinforcement Learning (RL) is then applied to further fine-tune the model, where an RL agent iteratively adjusts the parameters to improve the model fit using the R-square fitness parameter as the reward function to optimize. 

The rationale for RL among other machine learning algorithms was its ability as an open-ended adaptive optimization tool for dynamic treatment regimens and to fine-tune parameters of the GLV model for accurately predicting OVT dynamics\cite{yang2023reinforcement}. The RL's data-driven iterative learning in a delayed system adjusts parameters in response to changes in the GLV model behavior, which mimics adaptive responses in complex ecosystem interactions. Specifically, for the GLV equations, RL dynamically learns the predictive features of predator-prey interactions, accounting for time delays and non-linear feedback loops crucial in capturing oscillatory behaviors and dynamic patterns inherent in the ecosystem-based modeling of tumor-immune-virus interactions. In evidence to this rationale, \citet{pmlr-v85-yauney18a} demonstrated that RL agents optimize temozolomide dosing frequency by personalizing strategies to reduce tumor size and minimize toxicity, showing potential for improving oncolytic virotherapy by dynamically adjusting parameters based on patient-specific responses.

\subsection*{Data Collection and Preprocessing}

Experimental data were collected from cell viability assays and growth assays from our previous findings in \citep{Kiyokawa2021}. The cell viability data represented the population of virus-infected cells (predators), and the growth assay data represented the population of tumor cells (prey). These datasets were preprocessed by normalizing each dataset to the range \([0, 1]\) using min-max scaling. 

\subsection*{Datasets Descriptions}
\begin{itemize}
    \item \textbf{In Vitro Cell Killing Data:}
    \begin{itemize}
        \item \textit{Viability Assay:} Measures 005 (murine glioblastoma) cell death over 4 days, providing temporal dynamics of tumor cell viability under ICOVIR17 and combination therapies.
        \item \textit{Growth Assay:} Captures growth inhibition at 24, 72, and 120 hours, providing early and long-term response kinetics with metrics such as mean, standard deviation, and percentiles. Thus, early responses (24–72 hours) and long-term effects (up to 11 days) were modeled to predict tumor clearance and survival outcomes.
    \end{itemize}
    \item \textbf{Nanostring Immune Panel Data:} Gene expression profiles of glioblastoma samples treated with ICOVIR17 and anti-PD-1 sourced from \texttt{080119 PanCancer and Myeloid\_NormalizedData Kiyokawa.xlsx} . Comparative analyses (ICOVIR17 vs. Control, Combo vs. ICOVIR17, Combo vs. Anti-PD-1) reveal immune regulatory networks and key mediators.
\end{itemize}

The integrated GLV hybrid model combines tumor cell dynamics and immune regulatory insights, testing synergistic, additive, or antagonistic interactions against molecular expressions from animal survival data (Nanostring panel) to predict optimal therapeutic outcomes.

\subsection{Parameter Optimization}

\subsubsection{Genetic Algorithms (GA)}

The GLV model parameters were first optimized using Genetic Algorithms (GA), implemented using the \texttt{DEAP} library. The Genetic Algorithm (GA) was configured with a population size of 50 individuals and 40 generations. Crossover was performed using blend crossover (\texttt{tools.cxBlend}) with \( \alpha = 0.5 \) and a probability (\texttt{cxpb}) of 0.5, while mutation used Gaussian mutation (\texttt{tools.mutGaussian}) with \( \mu = 0 \), \( \sigma = 1 \), and a probability (\texttt{mutpb}) of 0.2. Selection was carried out using tournament selection with a tournament size of 3. The fitness function was defined as the sum of the mean squared error (MSE) between the predicted and experimental prey and predator populations. The best individual from the final population provided the optimal parameters.

\subsubsection{Differential Evolution (DE)}

After GA optimization, Differential Evolution (DE) was applied to refine the parameters further. The DE algorithm's optimization parameters, implemented using \texttt{scipy.optimize.differential\_evolution}, was configured  with bounds set to \([-1, 2]\), using the default strategy \texttt{best1bin}, a mutation factor (\(F\)) of 0.8, and a crossover probability (\(CR\)) of 0.7. The DE algorithm iteratively adjusted the GLV parameters to minimize the fitness function, yielding a refined set of parameters.

\subsubsection{Reinforcement Learning (RL)}

To further fine-tune the GLV model, a reinforcement learning (RL) approach was applied using the \texttt{Stable-Baselines3} library, specifically the Proximal Policy Optimization (PPO) algorithm. PPO optimizes a policy, a mapping between the RL agent's actions and states (from the data) by balancing exploration and exploitation. RL was chosen as it enables an adaptive agent to iteratively optimize model parameters by interacting with the environment (i.e., the action-state space), simulating tumor-virus dynamics. The reward function minimizes the prediction error (e.g., MSE) between model outputs (actions) and experimental data (state space), driving the agent to maximize model accuracy and predictive performance. The RL environment was configured with a continuous 4-dimensional action space representing adjustments to \(\text{damping\_x}\), \(\text{damping\_y}\), \(\text{growth\_mod\_x}\), and \(\text{growth\_mod\_y}\). The observation space consisted of the final predicted values of prey and predator populations, with a reward function defined as the negative MSE between predicted and experimental data. The learning rate was set to \(3 \times 10^{-4}\), and the total timesteps were 10,000. The PPO agent iteratively adjusted the parameters to maximize the reward, leading to a final set of parameters that best fit the experimental data.

\subsection{Feature Importance Analysis}

Post-hyperparameter optimization, a feature importance analysis was conducted on the Nanostring gene expression data, using Random Forests (RF). This analysis identifies key predictive genes influencing the oscillatory dynamics of tumor-virus interactions, providing insights into potential therapeutic vulnerabilities or biomarkers for precision medicine. 

\subsubsection{Random Forests}
A Random Forest Regressor was used to identify key genes influencing the dynamics of the tumor-virus interactions. The Random Forest model was configured with 100 estimators and no maximum depth, allowing nodes to expand until all leaves were pure or contained fewer than 2 samples. A random state of 42 was used, and 20\% of the dataset was allocated for testing, following an 80-20 train-test split.

The feature importances were calculated based on the decrease in prediction error for each feature (gene). Feature importance serve as a key tool in explainable AI, offering clinically meaningful insights to design precision biomarkers and guide targeted therapies in patient care. 

\subsection{Gene Set Enrichment}
g:Profiler was used for gene ontology and gene set enrichment analysis, identifying functional terms, pathways, and regulatory elements associated with the top 30 feature importance gene list. It uses a hypergeometric test to calculate the statistical significance of enrichment for each term. By default, g:Profiler applies the g:SCS multiple testing correction algorithm, which accounts for the hierarchical structure of gene ontologies, ensuring robustness and low false discovery.


\section {Results}

\subsection{The AI-powered Hybrid GLV model is a robust and accurate predictive algorithm for forecasting OVT-tumor dynamics.}

The hybrid model effectively learned the optimal parameters to fit the dynamics of the system. The optimized parameters in Table \ref{table:optimized_parameters} capture the predator-prey dynamics. The negative values of \(a\) and \(d\) represent natural declines in prey (tumor cells) and predator (virus-infected cells) populations, while negative \(b\) and \(c\) indicate antagonistic effects where predators infect prey, and prey depletion incurs a cost to predators. A delay of 0.3 reflects a time lag in predator-prey interactions, reflecting biological processes like immune responses or virus propagation.

The positive \(K_x\) suggests a carrying capacity for prey, while negative \(K_y\) implies the predator population is constrained or unstable without prey. Damping parameters (\(damping\_x\), \(damping\_y\)) indicate how quickly oscillations in prey and predator populations stabilize, while growth modifiers (\(growth\_mod\_x\), \(growth\_mod\_y\)) reflect the net growth trends under specific conditions. The parameters collectively minimize prediction error, modeling tumor (prey) and virus-infected cell (predator) interactions accurately. Table \ref{table:training_metrics} indicates that the PPO RL agent achieves stable training with a low policy gradient loss (1.21e-10) and effective exploration-exploitation balance (entropy loss: -5.68) while maintaining efficiency at 73 fps over 8192 timesteps.

\begin{figure}[h!]
    \centering
    \makebox[\textwidth]{ 
        \includegraphics[width=1.2\textwidth]{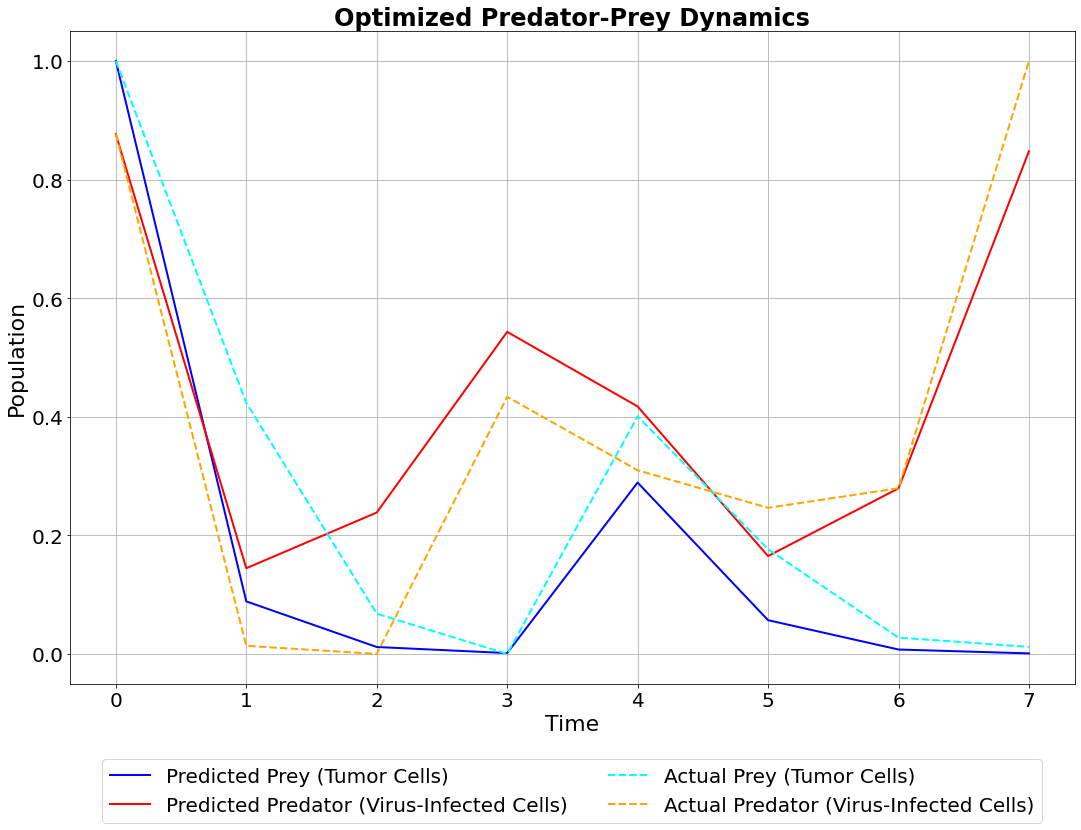}
    }
    \caption{Comparison of predicted and actual predator-prey dynamics using the optimized hybrid GLV model with damping and growth adjustments. Solid lines represent predicted oscillatory dynamics of prey (tumor cells) and predator (virus-infected cells), while dashed lines show experimental data-driven populations.}
    \label{fig:final_optimized_vs_experimental}
\end{figure}

As shown in Table~\ref{table:model_metrics}, the model demonstrates robustness by effectively capturing complex oscillatory dynamics, through empirical validation, by a low MSE (mean square error) and high R-square values for the fitness between the actual data and model predictions. Further, the use of innovative optimization algorithms, including reinforcement learning, genetic algorithms, and differential evolution, mimic real-world biological processes to optimize the timing (i.e., time-delays in tumor-virus interactions), dosage, and parameter tuning in OVT dynamics. The predictive power of these algorithms captured the oscillatory patterns in the combination therapy dynamics, as revealed in Figure~\ref{fig:final_optimized_vs_experimental}. These methods allow adaptive adjustments to clinically relevant therapy parameters, minimizing error and improving predictive accuracy for precision medicine.

\begin{table}[h!]
\centering
\begin{tabular}{|c|c|}
\hline
\textbf{Parameter} & \textbf{Optimized Value (GA)} \\ \hline
\textbf{a} & -0.966 \\ \hline
\textbf{b} & -0.087 \\ \hline
\textbf{c} & -0.557 \\ \hline
\textbf{d} & -0.331 \\ \hline
\textbf{delay} & 0.300 \\ \hline
\textbf{Kx} & 0.519 \\ \hline
\textbf{Ky} & -0.214 \\ \hline
\textbf{damping\_x} & 1.162 \\ \hline
\textbf{damping\_y} & 0.667 \\ \hline
\textbf{growth\_mod\_x} & -0.890 \\ \hline
\textbf{growth\_mod\_y} & 1.081 \\ \hline
\end{tabular}
\caption{Optimized parameters for the Generalized Lotka-Volterra (GLV) model using Genetic Algorithms (GA). These parameters have been fine-tuned to minimize the error between the model predictions and experimental data, capturing the dynamics of tumor (prey) and virus-infected cells (predator).}
\label{table:optimized_parameters}
\end{table}

\begin{table}[h!]
\centering
\begin{tabular}{|c|c|}
\hline
\textbf{Metric} & \textbf{Value} \\ \hline
\multicolumn{2}{|c|}{\textbf{Time}} \\ \hline
fps & 73 \\ \hline
iterations & 4 \\ \hline
time\_elapsed & 111 \\ \hline
total\_timesteps & 8192 \\ \hline
\multicolumn{2}{|c|}{\textbf{Training}} \\ \hline
approx\_kl & 0.0 \\ \hline
clip\_fraction & 0 \\ \hline
clip\_range & 0.2 \\ \hline
entropy\_loss & -5.68 \\ \hline
learning\_rate & 0.0003 \\ \hline
n\_updates & 30 \\ \hline
policy\_gradient\_loss & 1.21e-10 \\ \hline
std & 1 \\ \hline
\end{tabular}
\caption{Training metrics for the reinforcement learning (RL)(i.e., PPO an iterative and adaptive learning model). The fps (frames per second) indicates the speed of training, while iterations and total timesteps reflect the training duration. clip range, entropy loss, and policy gradient loss regulate the stability of updates, prevent over-fitting, and balance the exploration-exploitation tradeoff.}
\label{table:training_metrics}
\end{table}
\clearpage

\begin{table}[h!]
\centering
\begin{tabular}{|c|c|}
\hline
\textbf{Metric} & \textbf{Value} \\ \hline
MSE (Prey) & 0.0178 \\ \hline
MSE (Predator) & 0.0159 \\ \hline
$R^2$ Score (Prey) & 0.826 \\ \hline
$R^2$ Score (Predator) & 0.865 \\ \hline
\end{tabular}
\caption{Model metrics for the time-delay hybrid GLV model optimization. The low mean squared error (MSE) (loss function) values indicate that the model predictions are very close to the actual experimental data, while the high $R^2$ scores demonstrate a strong fit of the model to the complex data. These metrics are meaningful given the complexity of the system, which involves non-linear interactions, delays, and multiple parameters. The robustness of the model is demonstrated by its ability to accurately predict the ecological predator-prey dynamics in OVT trajectories.}
\label{table:model_metrics}
\end{table} 

\clearpage
\newpage
\begin{table}[h!]
\centering
\caption{Top 30 Importance Features extracted by Random Forests Using GLV Lotka-Volterra Equations Fitting on Nanostring Data for Combined Samples (Combination Therapy - ICOVIR17 and anti-PD1).}

\begin{tabular}{|l|l|}
\hline
\textbf{Gene} & \textbf{Importance} \\ \hline
Bid     & 0.023018 \\ \hline
Pdgfra  & 0.017200 \\ \hline
Polr2a  & 0.017048 \\ \hline
Tfrc    & 0.017034 \\ \hline
Oasl1   & 0.016955 \\ \hline
Isg20   & 0.016144 \\ \hline
Dusp2   & 0.016101 \\ \hline
Cd81    & 0.015827 \\ \hline
Traf2   & 0.015808 \\ \hline
Icam2   & 0.015656 \\ \hline
Il13ra1 & 0.015382 \\ \hline
Znrf2   & 0.015371 \\ \hline
Pecam1  & 0.014914 \\ \hline
Arhgef28& 0.013807 \\ \hline
Epcam   & 0.010119 \\ \hline
Ifit2   & 0.009548 \\ \hline
Prkci   & 0.009419 \\ \hline
Serinc2 & 0.009284 \\ \hline
Il7r    & 0.009282 \\ \hline
Gpi1    & 0.009114 \\ \hline
Adcyap1r1 & 0.009041 \\ \hline
Cfp     & 0.008989 \\ \hline
Angpt1  & 0.008951 \\ \hline
Klk1    & 0.008843 \\ \hline
Il18    & 0.008743 \\ \hline
Ets1    & 0.008732 \\ \hline
Il6ra   & 0.008725 \\ \hline
Prf1    & 0.008685 \\ \hline
Lgals3  & 0.008649 \\ \hline
Mmp12   & 0.008640 \\ \hline
\end{tabular}
\label{table:topmarkers}
\end{table}

\clearpage
\newpage
\subsection{Feature Mapping-based Biomarker Discovery Strongly Aligns with Molecular Findings from Previous Studies in a Data and Experiment-Agnostic Manner: Functional Validation of T-cells and TNF/NF-kB.}
As shown in Table~\ref{table:topmarkers}, the discovery of biomarkers, particularly related to immune response pathways like T cell activation and TNF/NF-kB, aligns with molecular signatures identified in prior studies, supporting functional validation and improving therapeutic targeting. Despite using only 10 importance features out of over 1000 genes from the Nanostring dataset, the model's agnostic approach identified statistically significant pathways matching prior results. 

\begin{figure}[ht]
\centering
\hspace*{-3.5cm}
\includegraphics[width=19cm]
{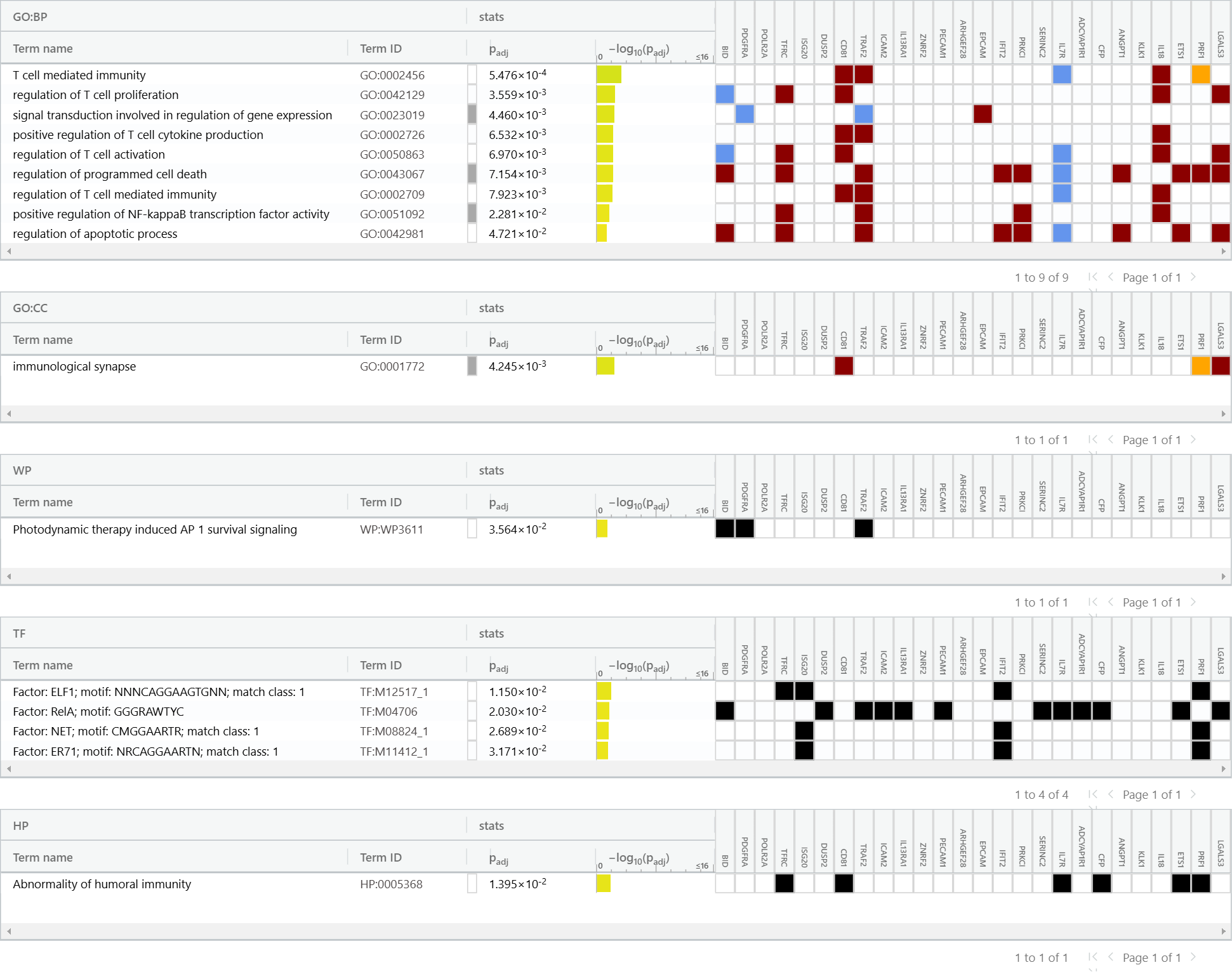} 
\caption{G:profiler visualization of Top 30 Importance Features for Predator-Prey Modelling on Nanostring Data. The color codes represent the source of evidence for the functional annotation. Computational evidence is shown in blue, while evidence generated by direct experiments is shown in red or orange. Black indicates the binary presence/absence of genes in the set.}
\label{fig:top30}
\end{figure}

\subsection{Photodynamic Therapy Activates Similar Immune and Signaling Mechanisms as Combined Therapy of Immune Checkpoint Inhibitor and OVT}

As shown in Table~\ref{table:enrichedfeatures} and Figure~\ref{fig:top30}, the RF-predicted salience map highlighted the most enriched features and their corresponding rankings. The model predicts that photodynamic therapy synergizes with immune checkpoint inhibitors and OVT by activating comparable immune and signaling mechanisms, such as T-cell-mediated responses and TNF/NF-KB pathways. Interestingly, our findings suggest that combination therapy of oncolytic virotherapy (ICOVIR17), a hyaluronidase-expressing adenovirus, combined with anti-PD-1 immune checkpoint blockade, elicits immune responses similar to those observed with photodynamic therapy. As seen in the gProfiler gene set enrichment analysis:

\textbf{Photodynamic Therapy AP-1 Pathway:} The identification of \textit{"Photodynamic therapy induced AP-1 survival signaling"} (WP:WP3611) with genes such as \textbf{BID}, \textbf{PDGFRA}, and \textbf{TRAF2} highlights the model's relevance and predictive power to the therapy context. This pathway is critical for cell survival and apoptosis regulation in response to therapy progression.

\textbf{Immune Responses:} The top hits, including \textit{"positive regulation of lymphocyte-mediated immunity"} and \textit{"positive regulation of leukocyte-mediated immunity,"} highlight pathways involving T cell regulation and leukocytes. Genes such as \textbf{TFRC} (iron regulation in T cells), \textbf{CD81} (immune modulation), and \textbf{TRAF2} (involved in NF-kB signaling and cell survival) are essential for immune response activation and regulation.

\textbf{Functional Validity:} The statistical significance (adjusted p-values ranging from 0.0333 to 0.0013) confirms the robustness of these pathways in the combination therapy model. The results demonstrate how the model can extract meaningful biological insights while remaining agnostic to input data.

Other signature genes such as CD81, IL7R, IL18, PRF1, and TFRC were identified as significant predictors of T cell regulation by G: profiler, particularly influencing T cell activation, proliferation, and cytokine production. These genes likely indicate activation of cytotoxic CD8+ T cells, given their roles in immune regulation and effector function.

Positive regulation of the NF-kappaB pathway was associated with genes such as TFRC, TRAF2, PRKCI, and IL18, emphasizing their role in promoting inflammatory and survival signals in T cells. Gene signatures like BID, TFRC, TRAF2, and IFIT2 predicted involvement in apoptotic processes, providing a balance between growth and decay in these predator-prey feedback loops. Furthermore, 
IL18 and NF-kappaB are key to neuro-immune and brain-gut-immune axes, with IL18 driving interferon-gamma signaling critical for OVT. IFN-gamma secretion could enhance antigen presentation, and drive Th1-mediated immunity in targeted OVT. A long-short term memory network (LSTM)-based feature importance analysis on the parameter-optimized GLV model further validated these findings, highlighting contributions from TNF signaling, specifically TNFR1, interleukins like IL19 and IL17, NF-kB signaling, extracellular matrix remodeling via MMP9 metalloproteinases, and the mTOR pathway (See Supplementary Information, Table-\ref{table:lstm_feature_importance}).

\begin{table}[h!]
\centering
\caption{Top Features Highlighting Functional Validity of the Model in Predicting Combined OVT Outcomes}
\begin{tabular}{|l|l|l|}
\hline
\textbf{Term Name} & \textbf{Adj. P.} & \textbf{Intersecting Genes} \\ \hline
Tumor necrosis factor receptor superfamily binding & 0.0333 & BID, TRAF2 \\ \hline
Positive regulation of lymphocyte mediated immunity & 0.0186 & TFRC, CD81, TRAF2 \\ \hline
Positive regulation of leukocyte mediated immunity & 0.0303 & TFRC, CD81, TRAF2 \\ \hline
Photodynamic therapy induced AP-1 survival signaling & 0.0013 & BID, PDGFRA, TRAF2 \\ \hline
\end{tabular}
\label{table:enrichedfeatures}
\end{table}

\newpage
\clearpage  

\section{Discussion}

\subsection{Computational Medicine: Predictive Models in Precision Oncology}

This interdisciplinary study integrates AI techniques such as RL, genetic algorithms GA, DE, and feature importance analysis via RF to forecast, optimize, and interpret complex OVT dynamics. By utilizing data-driven, explainable AI approaches, including salience maps (feature importance), the hybrid ecological model enhances predictions of OVT dynamics, for advancing precision oncology and systems medicine. These models, paired with experimental and patient-specific data, offer the ability to predict therapy responses. As shown in Figure~\ref{fig:final_optimized_vs_experimental}, the model forecasts the complex oscillatory behaviors of multi-agent interactions and identifies critical biomarkers for personalized treatment. The model provides a good estimate of the ecological (population) dynamics, closely matching the amplitude and frequency of oscillations.

\subsection{Shared Molecular Mechanisms Between Photodynamic Therapy and Combination Therapy Models}

As shown in Table 4, the GLV model predicted photodynamic therapy signatures as its top 10 importance features using random forests for the ICOVIR17 combination model's experimental data fit. The overlap of photodynamic therapy-induced AP-1 survival signaling pathways with the mechanisms identified for the combination therapy suggests shared molecular response mechanisms underlying these approaches to therapy. This highlights the role of critical dynamics between decay and growth processes, since signals of apoptosis regulation (BID, TRAF2) and cell survival or growth pathways (PDGFRA), are observed in the shared context. Such overlap indicates that targeting these pathways could optimize therapeutic efficacy across multiple treatment modalities, and fine-tune the oscillations of the predator-prey ecosystem. Further, it emphasizes their importance as predictive biomarkers and potential intervention points.

\subsection{Validating Hybrid GLV Model Predictions with Immune Activation Markers and Predator-Prey Dynamics}

Our hybrid GLV model predictions strongly align with the findings by \citet{Kiyokawa2021}, particularly the identification of T cell activation markers, TNF and NF-kB as predictive mediators. The model independently highlighted these features as key regulators of the tumor-virus-immune interactions. This agreement validates the computational model, supporting its ability to accurately identify critical regulators of the therapeutic processes and their complex dynamics in a context-agnostic manner, bridging in vitro cell dynamics and in vivo immune responses. As seen in Table 4, TRAF2 emerged as one of the top signals in salience mapping (Feature importance). TRAF2 plays a key role in TNF signaling, linking it to NF-kB activation, a central mediator of inflammation, immune responses, and cell survival (Table~\ref{table:enrichedfeatures}) \citep{borghi2016traf2}.  BID (apoptosis regulator) and PDGFRA (growth signaling) complement these processes by balancing cell death and proliferation, highlighting their relevance in OVT's oscillatory dynamics. Thus, these gene expression signatures themselves support the oscillatory dynamics of the model, wherein the OVT-immune-cancer interactions are in a predator-prey feedback loop, causing the tumor to adaptively oscillate between growth and decay, toward a 'stable attractor' (e.g., a limit cycle) during the therapy progression. Thus, an intuitive and explainable model like the hybrid GLV by capturing predator-prey dynamics, reflects the interplay of immune activation and suppression among viral and tumor interactions, as shown by its alignment with the Nanostring markers, enhancing biological interpretability.

None of the current state-of-the-art models have achieved the integration of predictive biomarkers as features with a strong statistical fit, demonstrated by the presented model's residual variance between the predicted and actual values (i.e., low MSE of 0.02) and an R\(^2\) value of 0.83 and 0.87 for the prey and predator, respectively (Table~\ref{table:model_metrics}). This highlights the hybrid GLV model's ability to accurately capture key biological dynamics while maintaining predictive robustness as an explainable AI system, marking a significant advancement in modeling therapy responses in computational medicine. 

\section{Limitations and Future Directions}

\subsection{Longitudinal Monitoring: Algorithms for Time-Series Data Integration}

By integrating multimodal and real-time (longitudinal) data into these predictive models, we enabled dynamic fine-tuning of OVT dosing regimens, potentially leading to optimized therapeutic efficacy, reduced side effects, and improved patient care.  The multimodal features can be further enhanced in the prospective model scaling with longitudinal tumor-immune markers (such as from multiomics, and liquid biopsies or blood profiles), to adaptively adjust therapy parameters.  By integrating continuous feedback mechanisms, clinicians can use model outputs to customize treatment plans dynamically, accounting for each patient’s unique tumor microenvironment. Furthermore, the revealed molecular patterns from the gene expression studies can be extended to time-resolved, longitudinal RNA-seq and multiomics in future studies. The identified patterns and gene signatures provide potential targets for therapeutic vulnerabilities and biomarkers of OVT progression.

Given the suitability of the GLV framework for longitudinal data forecasting, it could benefit from time-delay embedding attractor reconstruction and recurrent neural networks (RNNs) such as long short-term memory networks (LSTMs) to predict patterns and temporal dependencies\citep{huang2020detecting}. These approaches could refine predictions and enable the identification of subtle patterns and multimodal features in vivo or patient-specific responses to OVT, with more time-resolved data. In evidence, we used LSTM as a feature importance mapping method akin to RF and found that it validated our RF-based salient features, intersecting with markers regulating T-helper cell differentiation, extracellular matrix remodeling factors, and most signals being part of the TNF and NF-kB pathways, as shown in the supplementary information in in Tables \ref{table:lstm_top_genes} and \ref{table:lstm_feature_importance}.

This study does not fully explore how the observed oscillatory patterns and complex dynamics could directly influence tumor-specific outcomes in optimizing combination OVT. A deeper analysis of how these dynamics correlate with clinical metrics, such as in vivo model tumor progression, immune response, and viral persistence, would enhance the translational relevance of the preclinical model in paving dynamic treatment regimens. 

\subsection{Attractor Reconstruction: Stability, Bifurcations, and Critical Transitions}

However, the model incorporates temporal features adaptable for personalized therapies, such as (patient-specific) time delays, and carrying capacities, which could further strengthen the predictive power for precision oncology. Various dynamical systems approaches can be exploited using these temporal features to enhance the model \citep{mcgehee2008bifurcations, crutchfield1982fluctuations}. For instance, attractor reconstruction uses delay-embedded trajectories to analyze the long-term behavior and stability of systems. By embedding time-delay dynamics into phase space, it becomes possible to visualize attractor geometries steering the ecological dynamics, and identify limit cycles or their critical transitions to strange attractors\citep{shin2023critical}. Thereby, attractor reconstruction can help detect the stability and bifurcations of these oscillatory systems\citep{evers2024early, chisholm2009critical, mcgehee2008bifurcations, crutchfield1982fluctuations}. This enables enhanced therapeutic predictions and real-time monitoring. For instance, the time-delay parameters and growth rates predicted by the model provide explainable features in the temporal dynamics of tumor-virus interactions, such as the lag between viral infection and tumor response\citep{shin2023critical}. These parameters can guide the timing and frequency of viral dosing or combination therapy, ensuring interventions align with the predicted peak effectiveness. By tailoring these inputs to the patient-specific tumor growth-decay patterns, clinicians can potentially optimize therapeutic efficacy while minimizing side effects.

\subsection*{Conclusion}
This integrated hybrid model combines GLV equations with AI-driven optimization techniques, including RL, GA, DE, and feature importance analysis via random forest (RF), to model OVT dynamics, fine-tune parameters, and identify critical genes predicting complex tumor-immune interactions. The predator-prey oscillatory dynamics captured by the model could help inform adaptive treatment schedules and personalized strategies by adjusting therapy timing (frequency) with tumor behavior, immune response, and viral interactions, while patient-specific data can further optimize therapy. In specific, our hybrid GLV model forecasts T cell activation markers and cytokine signals, such as TNF and NF-kB pathways, identified by \citet{Kiyokawa2021}, as key mediators of OVT-immune-tumor dynamics. 

Biosignatures, including CD81, TRAF2, IL18, and BID as predicted by RF salience maps, and TNF, IL4, MMP9, MMP12, among other tumor microenvironment remodeling pathways identified by LSTM feature extraction (See Supplementary Information; Table~\ref{table:lstm_top_genes}), can guide personalized cancer immunotherapies by targeting immune checkpoint regulators, modulating T-cell differentiation, and influencing apoptosis. These markers also suggest targeting brain-immune axis inflammatory signals, supporting the development of immunomodulatory drugs. Thus, our proof-of-concept model provides a promising platform for real-time, longitudinal monitoring, adaptive control, and optimization of combination OVT in precision medicine, with predictive biomarkers informing novel targeted therapies to improve quality patient care and outcomes.

\section*{Data and Code Availability}
All codes are provided in the following GitHub Repository:
https://github.com/Abicumaran/OVT

\section*{Declarations}
The authors declare no competing interests.

\bibliographystyle{plainnat} 
\bibliography{references}
\newpage
\section{Supplementary Information}

\subsection*{Top 30 Importance Features for Predator-Prey Modelling on Nanostring Data}

\begin{table}[ht]
\centering
\hspace*{-1cm}
\begin{tabular}{|p{6cm}|p{3cm}|p{6cm}|}
\hline
\textbf{Term Name} & \textbf{P.adj} & \textbf{Intersections} \\ \hline
T cell mediated immunity & 0.000548 & CD81, TRAF2, IL7R, IL18, PRF1 \\ \hline
Regulation of T cell proliferation & 0.003559 & BID, TFRC, CD81, IL18, LGALS3 \\ \hline
Signal transduction involved in regulation of gene expression & 0.004460 & PDGFRA, TRAF2, EPCAM \\ \hline
Positive regulation of T cell cytokine production & 0.006532 & CD81, TRAF2, IL18 \\ \hline
Regulation of T cell activation & 0.006970 & BID, TFRC, CD81, IL7R, IL18, LGALS3 \\ \hline
Regulation of programmed cell death & 0.007154 & BID, TFRC, TRAF2, IFIT2, PRKCI, IL7R, ANGPT1, ETS1, PRF1, LGALS3 \\ \hline
Regulation of T cell mediated immunity & 0.007923 & CD81, TRAF2, IL7R, IL18 \\ \hline
Positive regulation of NF-kappaB transcription factor activity & 0.022811 & TFRC, TRAF2, PRKCI, IL18 \\ \hline
Regulation of apoptotic process & 0.047208 & BID, TFRC, TRAF2, IFIT2, PRKCI, IL7R, ANGPT1, ETS1, LGALS3 \\ \hline
Immunological synapse & 0.004245 & CD81, PRF1, LGALS3 \\ \hline
Photodynamic therapy induced AP 1 survival signaling & 0.035645 & BID, PDGFRA, TRAF2 \\ \hline
Factor: ELF1 & 0.011499 & TFRC, ISG20, IFIT2, PRF1 \\ \hline
Factor: RelA & 0.020303 & BID, DUSP2, TRAF2, ICAM2, IL13RA1, PECAM1, SERINC2, IL7R, ADCYAP1R1, CFP, ETS1, LGALS3 \\ \hline
Factor: NET & 0.026888 & ISG20, IFIT2, PRF1 \\ \hline
Factor: ER71 & 0.031714 & ISG20, IFIT2, PRF1 \\ \hline
Abnormality of humoral immunity & 0.013951 & TFRC, CD81, IL7R, CFP, ETS1, PRF1 \\ \hline
\end{tabular}
\caption{Top 30 Importance Features for Predator-Prey Modelling on Nanostring Data. The table shows immune pathways, signals, and transcription factors involved in T-cell regulation, cytokine production, apoptosis, and immune regulation. These features can inform the development of immunomodulatory drugs or precision immunotherapies by targeting specific pathways (e.g., NF-kappaB activity, T-cell proliferation) to enhance or suppress immune responses effectively.}
\label{table:features}
\end{table}

\subsection*{\textbf{LSTM (Feature Importance Extraction)}}

The Long Short-Term Memory (LSTM) network extracts feature importance by its ability to learn temporal dependencies in the data. Gene expression data from the Nanostring panel was preprocessed by transposing the dataset to organize genes as columns and aggregating it to match the length of the target model output with an 80:20 training-testing split. The LSTM model had 2 layers (one LSTM layer with 50 units and one Dense layer) and used the Adam optimizer (learning rate = 0.001), with early stopping (patience = 10) and a batch size of 32. The LSTM was trained on the data to predict features of the optimized GLV model. The absolute mean of the learned weights in the LSTM layer was calculated as a proxy for feature importance.

The LSTM salience map is a potentially robust approach for feature importance due to its ability to capture long-range patterns in time-series data. However, given our small sample size, Random Forest (RF) feature predictions were prioritized for the main results, while LSTM insights still demonstrated predictive power in identifying functionally validated immune-inflammatory regulatory signals and ECM remodeling pathways. We propose their scalability and translatability to preclinical in vivo studies with larger datasets.

\begin{table}[h!]
\centering
\begin{tabular}{|l|c|}
\hline
\textbf{Gene} & \textbf{Importance} \\ \hline
Dusp1 & 0.0367 \\ \hline
Bcl2 & 0.0367 \\ \hline
Angpt2 & 0.0366 \\ \hline
Clec4n & 0.0365 \\ \hline
Mafb & 0.0363 \\ \hline
Ecsit & 0.0361 \\ \hline
Ttk & 0.0360 \\ \hline
Stat4 & 0.0360 \\ \hline
Il4 & 0.0358 \\ \hline
Rps6 & 0.0358 \\ \hline
Ell2 & 0.0358 \\ \hline
Mmp9 & 0.0358 \\ \hline
Il27 & 0.0358 \\ \hline
Cebpg & 0.0357 \\ \hline
Tnfaip3 & 0.0357 \\ \hline
Hspb2 & 0.0357 \\ \hline
Icosl & 0.0357 \\ \hline
Mmp12 & 0.0357 \\ \hline
Aoah & 0.0356 \\ \hline
Tspan8 & 0.0356 \\ \hline
Tnf & 0.0356 \\ \hline
Irak2 & 0.0355 \\ \hline
Lrp6 & 0.0355 \\ \hline
C8a & 0.0354 \\ \hline
Ikbke & 0.0354 \\ \hline
Ly6g & 0.0354 \\ \hline
Tnfrsf1a & 0.0353 \\ \hline
Adamts4 & 0.0352 \\ \hline
Tnfsf9 & 0.0352 \\ \hline
Ccl17 & 0.0352 \\ \hline
\end{tabular}
\caption{Top 30 genes predicted from LSTM-based feature importance on the optimized hybrid GLV model.}
\label{table:lstm_top_genes}
\end{table}

\begin{table}[h!]
\centering
\hspace*{-3cm} 
\begin{tabular}{|l|c|p{7cm}|}
\hline
\textbf{Term Name} & \textbf{P.adj} & \textbf{Intersections} \\ \hline
T cell differentiation & $3.46 \times 10^{-3}$ & BCL2, MAFB, STAT4, IL4, IL27, TNFSF9 \\ \hline
IL-17 signaling pathway & $7.71 \times 10^{-6}$ & IL4, MMP9, TNFAIP3, TNF, IKBKE, CCL17 \\ \hline
NF-kappa B signaling pathway & 0.01 & BCL2, TNFAIP3, TNF, TNFRSF1A \\ \hline
TNF signaling pathway & 0.01 & MMP9, TNFAIP3, TNF, TNFRSF1A \\ \hline
mTOR signaling pathway & 0.05 & RPS6, TNF, LRP6, TNFRSF1A \\ \hline
TNFR1-induced proapoptotic signaling & $6.14 \times 10^{-5}$ & TNFAIP3, TNF, IKBKE, TNFRSF1A \\ \hline
Regulation of TNFR1 signaling & $7.69 \times 10^{-4}$ & TNFAIP3, TNF, IKBKE, TNFRSF1A \\ \hline
Signaling by Interleukins & 0.001 & BCL2, STAT4, IL4, MMP9, IL27, TNF, \\ 
 & & IRAK2, TNFRSF1A \\ \hline
Cytokine Signaling in Immune system & 0.003 & BCL2, STAT4, IL4, MMP9, IL27, TNF, \\ 
 & & IRAK2, TNFRSF1A, TNFSF9 \\ \hline
TNFR1-induced NF-kappa-B signaling pathway & 0.01 & TNFAIP3, TNF, TNFRSF1A \\ \hline
IL 19 signaling pathway & 0.01 & IL4, TNF, TNFRSF1A \\ \hline
Matrix metalloproteinases & 0.01 & MMP9, MMP12, TNF \\ \hline
Photodynamic therapy induced NF kB survival signaling & 0.02 & MMP9, TNF, TNFRSF1A \\ \hline
\end{tabular}
\caption{Gene set enrichment from Top 30 LSTM-based feature importance signatures on optimized hybrid GLV model on Nanostring data.}
\label{table:lstm_feature_importance}
\end{table}

\begin{figure}[ht]
\centering
\hspace*{-3.5cm}
\includegraphics[width=19cm]
{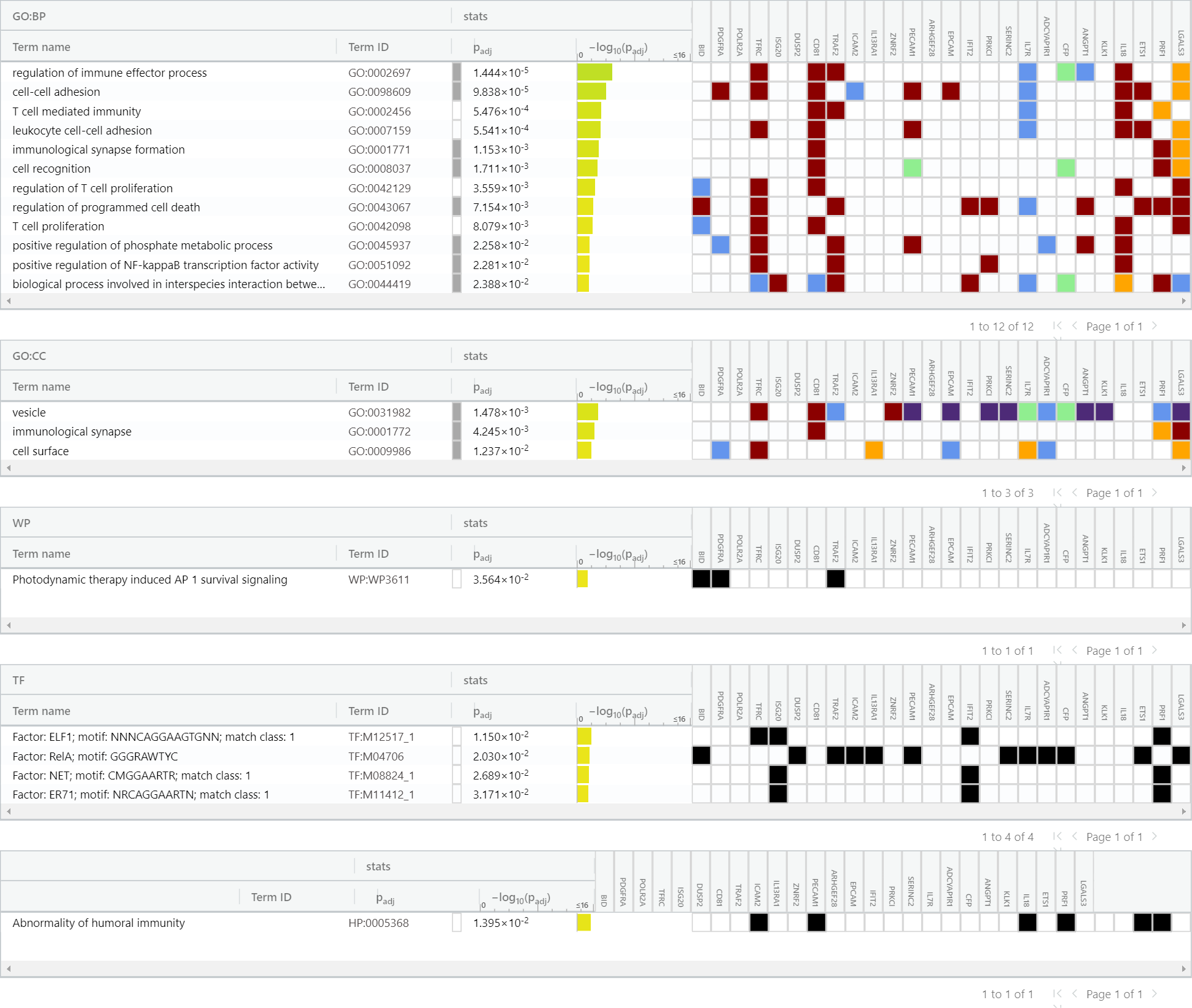} 
\caption{g:Profiler gene set enrichment for LSTM top 30 importance features predicted by the hybrid GLV model. The color codes represent the quality of the evidence for the functional annotation. Weaker evidence is depicted in blue, while strong evidence generated by direct experiments is shown in red or orange. Black indicates binary presence/absence of genes in the set.}
\label{fig:lstm_top30}
\end{figure}

\end{document}